\documentclass[twocolumn,showpacs,preprintnumbers,amsmath,amssymb]{revtex4}
\usepackage{graphicx}
\usepackage{dcolumn}
\usepackage{bm}
\begin{document}
\newcommand{\hs}{\hspace*{0.5cm}}
\newcommand{\vs}{\vspace*{0.5cm}}
\newcommand{\be}{\begin{equation}}
\newcommand{\ee}{\end{equation}}
\newcommand{\bea}{\begin{eqnarray}}
\newcommand{\eea}{\end{eqnarray}}
\newcommand{\ben}{\begin{enumerate}}
\newcommand{\een}{\end{enumerate}}
\newcommand{\nn}{\nonumber}
\newcommand{\crn}{\nonumber \\}
\newcommand{\non}{\nonumber}
\newcommand{\noi}{\noindent}
\newcommand{\al}{\alpha}
\newcommand{\la}{\lambda}
\newcommand{\bet}{\beta}
\newcommand{\ga}{\gamma}
\newcommand{\va}{\varphi}
\newcommand{\om}{\omega}
\newcommand{\pa}{\partial}
\newcommand{\fr}{\frac}
\newcommand{\bc}{\begin{center}}
\newcommand{\ec}{\end{center}}
\newcommand{\Ga}{\Gamma}
\newcommand{\de}{\delta}
\newcommand{\De}{\Delta}
\newcommand{\ep}{\epsilon}
\newcommand{\varep}{\varepsilon}
\newcommand{\ka}{\kappa}
\newcommand{\La}{\Lambda}
\newcommand{\si}{\sigma}
\newcommand{\Si}{\Sigma}
\newcommand{\ta}{\tau}
\newcommand{\up}{\upsilon}
\newcommand{\Up}{\Upsilon}
\newcommand{\ze}{\zeta}
\newcommand{\ps}{\psi}
\newcommand{\Ps}{\Psi}
\newcommand{\ph}{\phi}
\newcommand{\vph}{\varphi}
\newcommand{\Ph}{\Phi}
\newcommand{\Om}{\Omega}
\def\lappeq{\mathrel{\rlap{\raise.5ex\hbox{$<$}}
{\lower.5ex\hbox{$\sim$}}}}

\title{Surface Integral of Babu Diagram}

\author{Phung Van Dong}
 \email{pvdong@iop.vast.ac.vn}
\author{Hoang Ngoc Long}%
 \email{hnlong@iop.vast.ac.vn}
\affiliation{%
Institute of Physics, VAST, P. O. Box 429, Bo Ho, Hanoi 10000,
Vietnam }

\date{\today}

\begin{abstract}
We point out that the surface integral of the Babu diagram gives
non-trivial main contribution independent of the charged lepton
masses and the scale of new physics. It can  fit the experimental
data on both solar and atmospheric neutrino oscillations. However,
then the coupling constants of the new Higgs singlets with the
leptons gain values smaller than that in the usual two-loop
radiative analyses. This yields distinction between  our
consideration on ultraviolet-behaviors and the former based on the
infrared radiative corrections.
\end{abstract}


\maketitle

The evidence of neutrino oscillation  experimentally indicates
that neutrinos are massive particles and that flavor lepton number
is not conserved. Since in the standard model (SM), neutrinos are
massless and flavor lepton number is conserved. The neutrino
oscillation experiments are a clear sign that the SM has to be
extended.

There existed two different typical approaching ways on generating
neutrino masses such as the seesaw mechanism (SSM) \cite{seesaw}
and the radiative mechanism (RM) \cite{rad}. The SSM has been the
most popular explanation of the small neutrino masses. It is
simple, and relies only on dimensional analysis for the required
new physics. Current neutrino data point to a seesaw scale of
$M_R\sim 10^{10}-10^{15}\mbox{ GeV}$, where lepton number
violation occurs through the Majorana masses of the right-handed
neutrinos. With such a high scale, the effects of lepton flavor
violation in processes other than neutrino oscillation itself
become extremely small. For example, the branching ratio for the
decay $\mu\rightarrow e+\ga$ is of order $10^{-50}$ within the
seesaw extension of the SM.

An alternative to the SSM, the RM also explains the smallness of
neutrino masses naturally. In this approach, neutrino masses are
zero at the tree level and are induced only as finite radiative
corrections. These radiative corrections are {\it typically}
proportional to the square of the charged lepton (or quark) masses
divided by the scale of new physics. In this case, lepton flavor
violation in processes other than neutrino oscillations may become
experimentally accessible.

On RM, Babu has introduced a model \cite{babu} (called Babu model)
which is directly extended from the SM by thanking to two charged
singlet Higgs fields to induce neutrino masses at the two-loop
level. It has been judged that the neutrino masses arise to be
{\it naturally} small. However, if one takes to carefully
calculating that two-loop diagram (called Babu diagram), the one
will see it yields additionally a radiative mass term which is
{\it much} larger than the term derived by his comments and
calculations. Moreover, this term completely results from the
ultraviolet behaviors with large internal momenta of the diagram.
Further, if taking evaluation for neutrino masses and mixing, it
can  fit not only experiment data but also yields difference from
that based on the usual radiative corrections at low energy.

The aim of  this Letter is  to clear up these important remarks
and show some interesting consequences.

The Babu model \cite{babu} includes two $\mbox{SU}(2)_L$ singlet
Higgs fields, a singly charged field $h^+$ and a doubly charged
field $k^{++}$. Moreover, right-handed neutrinos are not
introduced. The addition of these singlets gives rise to the
Yukawa couplings: \be {\cal L}_Y =
f_{ab}\overline{(\psi_{aL})^C}\psi_{bL}h^+ +
h_{ab}\overline{(l_{aR})^C}l_{bR}k^{++}+h.c.,\label{e1} \ee where
$\psi_L$ stands for the left-handed lepton doublet, $l_R$ for the
right-handed charged lepton singlet and ($a$, $b$) being the
generation indices, a superscript $^C$ indicates charge
conjugation.  Here $\psi^C=C\overline{\psi}^T$ with $C$ is the
charge-conjugation matrix. The coupling constant $f_{ab}$ is
antisymmetric ($f_{ab}=-f_{ba}$), whereas $h_{ab}$ is symmetric
($h_{ab}=h_{ba}$). Gauge invariance precludes the singlet Higgs
fields from coupling to the quarks.  Eq. (\ref{e1}) conserves
lepton number, therefore, itself cannot be responsible for
neutrino mass generation. The Higgs potential contains the terms:
\be V(\phi,h^+,k^{++})=\mu(h^- h^- k^{++}+h^+ h^+
k^{--})+\cdot\cdot\cdot, \label{ee1}\ee which violate lepton
number by two units. They are expected to cause Majorana neutrino
masses.

\begin{figure}
\includegraphics{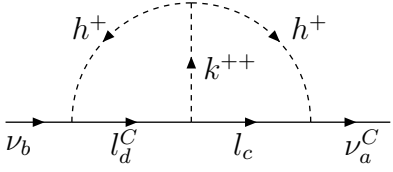}
\caption{\label{fig:fig1} The two-loop diagram in the Babu model.}
\end{figure}

On Babu's opinion in his model, Majorana neutrino masses are
generated at the two-loop level via the diagram shown in
\cite{babu} and again depicted in Fig.\ref{fig:fig1}. Then, he has
written down the corresponding mass matrix element for Majorana
neutrinos as follows

\bea M_{ab}&=&8\mu
f_{ac}\widetilde{h}_{cd}m_cm_dI_{cd}(f^+)_{db},\label{b1212}\eea
in which $\widetilde{h}_{ab}=\eta h_{ab}$ with $\eta=1$ for $a=b$
and $\eta=2$ for $a\neq b$. The integral $I_{cd}$ is given by \bea
I_{cd}&=&\int\fr{d^4k}{(2\pi)^4}\int\fr{d^4q}{(2\pi)^4}\fr{1}{k^2-m^2_c}
\fr{1}{k^2-m^2_h}\fr{1}{q^2-m^2_d}\crn
&\times&\fr{1}{q^2-m^2_h}\fr{1}{(k-q)^2-m^2_k}. \label{ba}\eea
After evaluating (\ref{ba}), it has been seen that the neutrino
masses are small finite and compatible with the experimental data
\cite{babu,babu1}.

However, now we will point out that the above solution of $M_{ab}$
is not complete. To see this, we will carefully apply the Feynman
rules to the diagram. After that, the large corrective term is
followed which kept in his model.

In the momentum space, the propagator of the charge-conjugate
spinor field is given as follows \cite{ch} \bea \langle
0|T(\psi^C\overline{\psi^C})|0\rangle &=& C\langle
0|T(\psi\overline{\psi})|0\rangle^T C^{-1} \crn \longrightarrow
CS^T(p)C^{-1}&=&(-p\!\!\!/-m)^{-1}=S(-p).\nn\eea For simplicity,
$f_{ab}$ and $h_{ab}$ are assumed to be real. Then, the mass
matrix of the Majorana neutrinos which is induced by the above
diagram {\it reads} \begin{widetext}\bea
-iM'_{ab}&=&i\sum_{c,d}(-8\mu f_{ac}\widetilde{h}_{cd}f_{db})\int
\fr{d^4 k}{(2\pi)^4} \fr{d^4 q}{(2\pi)^4}\fr{P_L(k\!\!\!\!/ +
m_c)P_L(-q\!\!\!/ + m_d)P_L}
{(k^2-m^2_c)(q^2-m^2_d)[(q-k)^2-m^2_k](k^2-m^2_h)
(q^2-m^2_h)}.\label{twoloop}\eea \end{widetext}  The $P_L$s {\it
in} the vertex factors have to be kept as coefficients {\it out}
of the partial momentum integrals defined for the free spinor and
boson fields. Even if their overall integral is obtained, they
still are in the coefficient. Keep our mind that, here the
(\ref{twoloop}) has {\it conveniently} been written.

We have \bea P_L(k\!\!\!\!/ + m_c)P_L(-q\!\!\!/ &+& m_d)P_L= \crn
&-& P_Lk\!\!\!\!/P_Lq\!\!\!/ P_L+ P_L k\!\!\!\!/ P_Lm_d \crn &-&
m_cP_Lq\!\!\!/P_L+ m_c m_d P_L.\label{tr}\eea Therefore, the r.h.s
of (\ref{twoloop}) is summed of four terms respectively to the
terms in the r.h.s of (\ref{tr}). The fourth is finite which has
been considered by Babu in his paper. Thus, applying its $P_L$ on
the left-handed Majorana field in the mass Lagrangian, e.g. $-\fr
1 2 \overline{(\nu_L)^C}M'^{(4)}P_L\nu_L=-\fr 1 2
\overline{(\nu_L)^C}M'^{(4)}\nu_L$, we get then: \bea
M'^{(4)}_{ab}=8\mu
\sum_{c,d}f_{ac}\widetilde{h}_{cd}f_{db}m_cm_dI_{cd},\label{uc1}\eea
with $I_{cd}$ given in (\ref{ba}). For the second and the third,
their integrals are finite, while the respective coefficients are
vanish. Thus they are explicitly canceled: \bea
M'^{(2)}_{ab}=M'^{(3)}_{ab}=0.\label{eq23}\eea

Finally, to keep the $P_L$s in the coefficient of the respective
overall momentum integral, the first term can be explicitly given
in the following form:\begin{widetext}\be
-iM'^{(1)}_{ab}=8i\mu\sum_{c,d}
f_{ac}\widetilde{h}_{cd}f_{db}P_L\ga_\mu P_L\ga_\nu P_L\int
\fr{d^4 k}{(2\pi)^4} \fr{d^4 q}{(2\pi)^4} \fr{k^\mu
q^\nu}{(k^2-m^2_c)(q^2-m^2_d)[(q-k)^2-m^2_k](k^2-m^2_h)
(q^2-m^2_h)}.\label{uc} \ee\end{widetext} It can be checked that
the two-loop integral of $k,q$ in the r.h.s of (\ref{uc}) is
logarithmically divergent while its coefficient is vanish, hence
\be -iM'^{(1)}_{ab}=0\times\infty,\label{38}\ee which follows an
{\it indeterminate} radiative term of the Babu diagram.

Explicitly, this term has been canceled in the former analyses
\cite{babu,babu1,ex} which present a fallacy. Suppose \bea P_L
k\!\!\!\!/ P_L q\!\!\!/ P_L &=& P_L P_R k\!\!\!\!/ q\!\!\!/ P_L,
\label{fall1} \eea happens in (\ref{twoloop}). Indeed, the source
of the fallacy is that algebraically permuting of the matrices
such as (\ref{fall1}) in the integrand of the logarithmically
divergent momentum integral of (\ref{twoloop}) changes the value
of the integral by an finite amount, which is equal to (\ref{38})
(also see \cite{chli}).

To evaluate (\ref{38}), dimensional regularization can be
developed. Since the anticommutativity of $\ga_5$ in $d$
dimensions is lost \cite{more} (also see \cite{th}), the idea is
that \cite{pes}: The external indices and the momenta all live in
the physical four dimensions, the loop momenta $k,q$ have
components in all dimensions. $\ga_5$ is expanded continuously
analytic to d-dimensions such that anti-commutes with $\ga_\mu$
for $\mu=0,1,2,3$ but commutes with $\ga_\mu$ for other values of
$\mu$. Let $k = k_{\parallel}+k_{\perp}, q =
q_{\parallel}+q_{\perp}$ we have then \cite{pes} \bea P_L
k\!\!\!\!/ P_L q\!\!\!/ P_L=P_L
k\!\!\!\!/_{\perp}q\!\!\!/_{\perp}\rightarrow
P_L\fr{d-4}{d}kq.\label{10}\eea The coefficient in (\ref{10}) owns
$d-4$ which yields zero-limit whenever returned back to the four
dimensions, hence we need only to define the divergent parts of
the integral in (\ref{uc}). By denoting $4-d=2\epsilon$, it can be
checked that\bea
I&\equiv&\int\fr{d^dk}{(2\pi)^d}\fr{d^dq}{(2\pi)^d}\fr{kq}{(k^2-m^2_c)
(q^2-m^2_d)[(q-k)^2-m^2_k]}\crn
&\times&\fr{1}{(k^2-m^2_h)(q^2-m^2_h)}
 =-\fr{1}{2(4\pi)^4\epsilon}+{\cal
O}(1).\eea Then, the indeterminate term yields \bea
0&\times&\infty = 8i\mu\sum_{c,d}
f_{ac}\widetilde{h}_{cd}f_{db}P_L\fr{d-4}{d}\times
I\nn\\&=&-4i\mu\sum_{c,d}
f_{ac}\widetilde{h}_{cd}f_{db}P_L\left(\epsilon+\fr{\epsilon^2}{2}\right)
 \left[-\fr{1}{2(4\pi)^4\epsilon}+{\cal O}(1)\right]\crn &=&
\fr{2i}{(4\pi)^4}\mu \sum_{c,d}f_{ac}\widetilde{h}_{cd}f_{db}P_L+
{\cal O}(\epsilon).\eea By shifting the $P_L$ into the left-hand
Majorana field in its mass Lagrangian, we can rewrite\bea
M'^{(1)}_{ab}&=& -\fr{2}{(4\pi)^4}\mu
\sum_{c,d}f_{ac}\widetilde{h}_{cd}f_{db}+{\cal
O}(\epsilon).\label{uc3}\eea Hence, with the help of (\ref{uc1}),
(\ref{eq23}) and (\ref{uc3}), the mass matrix of the Majorana
neutrinos in the four dimensions is given as follows \bea
M'_{ab}=&-&\fr{2}{(4\pi)^4}\mu
\sum_{c,d}f_{ac}\widetilde{h}_{cd}f_{db}\crn &+&8\mu
\sum_{c,d}f_{ac}\widetilde{h}_{cd}f_{db}m_cm_dI_{cd}.\label{tol}\eea

The following remarks are in order: \ben \item The radiative term
(\ref{uc3}) is dependent only on the Yukawa coupling constants of
new physics such as $f_{ab},h_{ab},\mu$, but not on the masses of
the charged leptons and of the new Higgs scalars. Thus, it should
{\it also} be present in the massless theory. \item It can be
checked that this term is {\it not} affected by higher-order
radiative corrections, e.g. the Babu diagram  with more than two
loops do not contribute to this term.
\item The term completely results from the ultraviolet asymptotic
(large momenta) of the integrand in (\ref{uc}) and thus in
(\ref{twoloop}), is also identified as the {\it surface} integral
of the diagram. Let us analyze that, the integrand in the
(\ref{uc1}) is strongly local, large around zero-momenta and hence
gives main distribution to this type of the RM, as usual (infrared
behavior); while the (\ref{uc3}) is distinct from that upon
ultraviolet behavior.
\item On the evaluation of the Babu integral $I_{cd}$ in \cite{babu}
(also see \cite{donald,babu1,ex}), the terms in (\ref{uc1}) are
proportional to, e.g. $m_cm_d/\Lambda$ with $\Lambda\sim
m^2_h,m^2_k$, therefore, it is very smaller than that in the
(\ref{uc3}). For example, choosing $m_k\simeq m_h \sim 100\
\mbox{GeV}$, the largest one is only $m^2_{\tau}/\Lambda \approx
0.00031 \ll 1$. Thus the surface integral gives {\it main}
contribution on the Babu diagram. The mass matrix (\ref{tol}) can
be approximated by \bea M'_{ab}&\simeq&-\fr{2}{(4\pi)^4}\mu
\sum_{c,d}f_{ac}\widetilde{h}_{cd}f_{db},\label{b22}\eea  which
completely differs from early that in (\ref{b1212}).\een

It can be checked that the matrix $M'_{ab}$ in (\ref{tol}) as well
as its approximation in (\ref{b22}) always has a zero eigenvalue
{\it independent} on its parameters; and the respective eigenstate
is dependent {\it only} on the $f_{ab}$ couplings, which
is identified as: \bea m_{\nu_1}&=&0,\\
\nu_1 &=& \fr{1}{\sqrt{f^2_{e\mu}+f^2_{e\tau}+f^2_{\mu\tau}}}
(f_{\mu\tau}\nu_e-f_{e\tau}\nu_\mu+f_{e\mu}\nu_\tau).\nn\\
\label{eigz}\eea Next, from the orthogonal condition of the
$\nu_1$ eigenstate to the two remaining neutrino vectors, one  can
write \bea \nu'_2 &=&
\fr{1}{\sqrt{f^2_{e\tau}+f^2_{\mu\tau}}}(f_{e\tau}\nu_e+f_{\mu\tau}\nu_{\mu}),\crn
\nu'_3 &=& \fr{1}{\sqrt{f^2_{e\mu}+f^2_{e\tau}}}(f_{e\mu}\nu_\mu
+f_{e\tau}\nu_{\tau}).\label{eigz1}\eea Then in this basis of
$(\nu_1,\nu'_2,\nu'_3)$, the matrix
$M'_{ab}$ in  (\ref{tol}) owns the form: \bea M''=\left(%
\begin{array}{ccc}
  0 & 0 & 0 \\
  0 & m_{\nu'_{2}} & m_{\nu'_{2}\nu'_{3}} \\
  0 & m_{\nu'_{2}\nu'_{3}} & m_{\nu'_{3}} \\
\end{array}%
\right).\label{mix}\eea The matrix $M''_{ab}$ gives mixing of
$\nu'_2$ and $\nu'_3$. Next, by rotating over the mixing angle
$\phi$ in the plane $(\nu'_2,\nu'_3)\rightarrow (\nu_2,\nu_3)$, we
have\bea \tan 2 \phi = \fr{2 m_{\nu'_2 \nu'_3}}{m_{\nu'_3
}-m_{\nu'_2 }},\eea and the physical mass eigenvalues are given by
\be m_{\nu_2}=\fr 1 2
\left[m_{\nu'_2}+m_{\nu'_3}-\sqrt{(m_{\nu'_2}-m_{\nu'_3})^2+4
m^2_{\nu'_2\nu'_3}}\right],\ee\be m_{\nu_3}=\fr 1 2
\left[m_{\nu'_2}+m_{\nu'_3}+\sqrt{(m_{\nu'_2}-m_{\nu'_3})^2+4
m^2_{\nu'_2\nu'_3}}\right]. \ee

With the help of (\ref{eigz}), (\ref{eigz1}) and the $\phi$, we
can easily obtain the mixing angles of the three neutrino
generations in this model respective to the standard
parametrization of an $3\times3$ Pontecorvo-Maki-Nakagawa-Sakata
(PMNS) \cite{ponte,alberico} unitary mixing matrix: \bea
\sin\theta_{13}&=&\fr{f_{e\tau}}{\sqrt{f^2_{e\tau}+f^2_{\mu
\tau}}}\sin\phi,\label{t13}\\
\tan\theta_{12}&=&
\fr{f_{e\tau}\sqrt{f^2_{e\mu}+f^2_{e\tau}+f^2_{\mu\tau}}}
{f_{\mu\tau}\sqrt{f^2_{e\tau}+f^2_{\mu\tau}}}\cos\phi,\label{t12}\\
\tan\theta_{23}&=&\fr{f_{\mu\tau}\sqrt{f^2_{e\mu}+f^2_{e\tau}}
\sin\phi+f_{e\mu}\sqrt{f^2_{e\tau}+f^2_{\mu\tau}}\cos\phi}
{f_{e\tau}\sqrt{f^2_{e\tau}+f^2_{\mu\tau}}\cos\phi}.\crn\label{t23}\eea
The equations (\ref{t13}), (\ref{t12}) and (\ref{t23}) yield that
the mixing angles $\theta_{12},\theta_{23}$ and $\theta_{13}$ can
be given by studying the $\phi$ on the mixing between $\nu'_2$ and
$\nu'_3$. Moreover, since the mixing angle $\theta_{13}$ is small,
e.g. the CHOOZ experiment gives $\sin^2\theta_{13} \preceq 0.08$
(also see \cite{alberico}), thus Eq. (\ref{t13}) requires that the
mixing angle $\phi$ is respective small too.

Now we will show that above solution can fit the current data on
both solar and atmospheric oscillations. For simplicity, we shall
work with the matrix (\ref{b22}) by taking a special case of
$\theta_{13}=0$. It yields $\phi=0$, and then:
\bea m_{\nu'_2\nu'_3}&=&0,\label{mmix}\\
\tan\theta_{12}&=&
\fr{f_{e\tau}\sqrt{f^2_{e\mu}+f^2_{e\tau}+f^2_{\mu\tau}}}
{f_{\mu\tau}\sqrt{f^2_{e\tau}+f^2_{\mu\tau}}},\label{t121}\\
\tan\theta_{23}&=&\fr{f_{e\mu}}{f_{e\tau}}.\label{t231}\eea Next,
on the constraints in the various parameters \cite{babu,babu1}, it
has been agreed that $h_{ab}\simeq 0$ for $a\neq b$, then with the
help of (\ref{b22}), (\ref{eigz}) and (\ref{eigz1}) the elements
of the matrix (\ref{mix}) yield \bea
m_{\nu'_2}&=&\fr{2\mu}{(4\pi)^4}\left[(f^2_{e\tau}+f^2_{\mu\tau})h_{\tau\tau}
\right.\crn &+& \left.\fr{f^2_{e\mu}}{f^2_{e\tau}+
f^2_{\mu\tau}}(f^2_{e\tau}h_{\mu\mu}
+f^2_{\mu\tau}h_{ee})\right],\label{m1}\\
m_{\nu'_3}&=&\fr{2\mu}{(4\pi)^4}\left[(f^2_{e\mu}+f^2_{e\tau})h_{ee}
\right.\crn &+&
\left.\fr{f^2_{\mu\tau}}{f^2_{e\mu}+f^2_{e\tau}}(f^2_{e\mu}h_{\tau\tau}
+f^2_{e\tau}h_{\mu\mu})\right],\label{m2}\\
m_{\nu'_2\nu'_3}&=&\fr{-2\mu f_{e\mu}f_{\mu\tau}}{(4\pi)^4
 \sqrt{(f^2_{e\mu}+f^2_{e\tau})
(f^2_{e\tau}+f^2_{\mu\tau})}}\crn &\times
&[f^2_{e\tau}(h_{\mu\mu}-h_{ee}-h_{\tau\tau})
-f^2_{e\mu}h_{ee}-f^2_{\mu\tau}h_{\tau\tau}].\crn \label{mm1}\eea
The equations (\ref{mmix}) and (\ref{mm1}) give \bea
h_{\mu\mu}=\fr{h_{ee}(f^2_{e\mu}+f^2_{e\tau})
+h_{\tau\tau}(f^2_{e\tau}+f^2_{\mu\tau})}{f^2_{e\tau}}.\label{ct}\eea
In addition, since $\nu'_2$ and $\nu'_3$ are not mixed, the
(\ref{m1}) and (\ref{m2}) yield \be
m_{\nu_2}=\fr{2\mu}{(4\pi)^4}(f^2_{e\mu}+f^2_{e\tau}+f^2_{\mu\tau})
\left(h_{\tau\tau}
+h_{ee}\fr{f^2_{e\mu}}{f^2_{e\tau}+f^2_{\mu\tau}}\right),\label{l1}\ee\be
m_{\nu_3}=\fr{2\mu}{(4\pi)^4}(f^2_{e\mu}+f^2_{e\tau}+f^2_{\mu\tau})\left(h_{ee}
+h_{\tau\tau}\fr{f^2_{\mu\tau}}{f^2_{e\mu}+f^2_{e\tau}}\right).\label{l2}\ee

Upon (\ref{t121}) and (\ref{t231}), we see that $\theta_{12}$ and
$\theta_{23}$ are not necessarily small. Then to fit with
observations (also see \cite{alberico}), one can choose: \bea
\tan^2 \theta_{23}\simeq1,\  \tan^2 \theta_{12}\simeq 0.43,\eea
which yield \bea f_{e\mu}\simeq f_{e\tau}\approx
0.585f_{\mu\tau}.\label{l3}\eea Denoting $\triangle
m^2_{ij}=m^2_{\nu_i}-m^2_{\nu_j}$, from experimental data
\cite{alberico} on solar and atmospheric neutrino oscillations,
one  gets the following relation $\triangle m^2_{21}/\triangle
m^2_{32}\approx 6.9\times 10^{-5} \mbox{ eV}^2/2.6\times 10^{-3}
\mbox{ eV}^2$. With the help of the (\ref{l3}), the equations
(\ref{l1}) and (\ref{l2}) give the following approximation:\bea
h_{ee}\approx -8.123 h_{\tau\tau},\label{l4}\eea which yields \bea
m_{\nu_2}&\approx&1.782
\times10^{-5}\mu f^2_{\mu\tau}h_{ee},\\
m_{\nu_{3}}&\approx&1.108\times10^{-4}\mu f^2_{\mu\tau}h_{ee}.\eea
Let us put $f_{\mu\tau}\simeq h_{ee}$ \cite{babu}, then it follows
\bea f_{\mu\tau}\simeq h_{ee}\approx \fr{7.754 \mbox{
eV}^{1/3}}{\mu^{1/3}}.\eea Choosing $\mu=200$ GeV, which is the
highest scale in the theory, we see that $f_{\mu\tau},h_{ee}
\approx 1.33\times10^{-3}$. On the constraints (\ref{ct}),
(\ref{l3}) and (\ref{l4}), the remaining
couplings gain: \bea f_{e\mu}&\simeq&f_{e\tau}\approx0.775\times10^{-3},\\
h_{\mu\mu}&\approx&2.01\times 10^{-3}, h_{\tau\tau}\approx
-1.63\times 10^{-4}.\eea It is easy to check that all above values
for the couplings explicitly satisfy the experimental constraints
given in \cite{babu,babu1}. Moreover, the above solution can be
also used to explain the mass hierarchy and mixing of Majorana
neutrinos at some energy scale wherein the gauge symmetry of the
SM vacuum is {\it not} broken.

Thus, we have derived a new {\it main} radiative term of the Babu
diagram which can fit the data on both solar and atmospheric
neutrino oscillations. However the required couplings $f,h$ for
the model are smaller than that in the usual two-loop radiative
analyses (also see \cite{aiza}). The reason lies behind
ultraviolet behavior on the radiative mechanism of generating
neutrino masses.

To conclude,  in this Letter we have shown that the Babu diagram
owns a non-trivial main term in ultraviolet-behavior. Its
contribution is given as a natural consequence of the surface
integral which is defined in the term of dimensional
regularization with the new properties of $\ga_5$ in
$d$-dimensions. After that, we have taken the exact
diagonalization of the neutrino mass matrix with the help of the
intermediate mixing angle $\phi$. Requirement of $\phi'$s
smallness  naturally gives the mixing parts  consistent with
current experimental data on both solar and atmospheric neutrino
oscillations. This consideration predicts the coupling constants
$f_{ab}$ and $h_{aa}$ are smaller than that in the former two-loop
analyses \cite{babu,babu1,ex,aiza,rad}.

It is also evident that the ultraviolet-asymptotic behavior in the
radiative mass mechanism should be mentioned in the current
neutrino problems. This conclusion adds one more nice feature to
the Babu model.

The authors wish to thank  T. Kitabayashi and  M.
Yasu$\grave{\mbox{e}}$ for reading the manuscript and comments

\end{document}